# Locality Based Core Selection for Multicore Shared Tree Multicasting

Koushik Sinha, Bidyut Gupta, Shahram Rahimi and Ashraf Alyanbaawi

Department of Computer Science
Southern Illinois University
Carbondale, IL, 62901, USA
(koushik.sinha, bidyut, shahram)@cs.siu.edu, ashraf@siu.edu

**Abstract**

Multicasting can be done in two different ways: source based tree approach and shared tree approach. Protocols such as Core Based Tree (CBT), Protocol Independent Multicasting Sparse Mode (PIM-SM) use shared tree approach. Shared tree approach is preferred over source-based tree approach because in the later construction of minimum cost tree per source is needed unlike a single shared tree in the former approach. In this paper, we present a candidate core selection approach for shared tree multicasting so that in a multicast session different senders can select different cores from the candidate core set based on the senders' physical locations to allow an efficient multicore multicasting approach.
**keywords**: core selection, pseudo diameter, multicore multicasting

## 1 Introduction

Multicast routing can be described as simultaneous delivery of data stream to multiple destinations. With the growth of internet, multicast routing has gained its importance in typical applications which include numerous forms of audiovisual conferencing and broadcasting, negotiation and e-commerce systems, replicated database querying, online games as well as the trivial resource discovery feature of Internet routers. Various multicast communication protocols have been developed, including flooding, spanning trees, reverse path forwarding, and core-based trees (CBT) [1], [2], [4], [6], [12] - [14].

In general, multicast communication protocols can be classified into two categories, namely, source-based trees [15], [1] and core based trees [2]. A problem associated with source-based-tree routing is that a router has to keep the pair information (source, group) and it is a one tree per source. In reality the Internet is a complex heterogeneous environment, which potentially has to support many thousands of active groups, each of which may be sparsely distributed; this technique clearly does not scale. Shared tree based approaches like CBT [2], [11] and protocol independent multicasting – sparse mode (PIM-SM) [6] offer an improvement in scalability by a factor of the number of active sources.

A core-based tree/shared tree [2] involves a single node, known as the core of the tree, from which branches emanate. These branches are made up of other routers, so-called non-core routers, which form a shortest path between a member-host's directly attached router and the core. A core need not be topologically centered, since multicasts vary in nature and therefore, the form of a core-based tree also can vary [2]. CBT is attractive compared to source based tree because of its key architectural features like scaling, tree creation, and unicast routing separation.

The major concerns of shared tree approach are: i) core selection and, ii) core as a single point failure. Core selection [7] is the problem of appropriate placement of a core or cores in a network for the purpose of improving the performance of the tree(s) constructed around these core(s) and thereby the performance of the multicast routing protocol according to some predetermined metrics.

In static networks core selection depends on knowledge of entire network topology. It involves all routers in the network. There exist several important works [3], [5], [13], [14] which take into account network topology while selecting a core. Maximum Path Count (MPC) core selection method [3] needs to know complete topology to calculate shortest paths for all pairs. The nodes are then sorted in descending order of their path counts. The first nodes are selected to be the candidate cores. In Delay Variant Multicast Algorithm (DVMA) it is assumed that the complete topology is available at each node [5]. It works on the principle of k-shortest paths to the group of destination nodes concerned. If these paths do not satisfy a delay constraint, then it may find a longer path, which is a shortfall of DVMA. Optimal Cost Based Tree (OCBT) [13], [14] approach calculates the cost of the tree rooted at each router in the network and selects the one which gives the lowest maximum delay over all other roots with lowest cost. It needs knowledge of the whole topology.

**Our Contribution**: In this work, we consider networks which use distance vector routing (DVR) protocol for communication, i.e. routers do not have knowledge of entire network topology. We shall use the concept of pseudo diameter [8-10] to approximate the idea used in OCBT to select the candidate cores such that the core with the lowest pseudo diameter will be the first to be selected, followed by the next lowest one and so on. Note that multiple cores always can be used to add an element of robustness. We will state a method of how during a multicast session a sender node (router) will determine which core will give the minimum delay considering the physical locations of itself and also the cores. The present work will select cores considering all routers in a network, yet does not need to know entire network topology unlike the OCBT approach.

The paper is organized as follows. In section 2 we will state briefly the concept of pseudo diameter. In Section 3 we will present the candidate core selection algorithm followed by an example. In Section 4 we will present the locality-based core selection approach Finally, Section 5 draws the conclusion.

## 2 Pseudo Diameter

Two widely used network routing protocols are distance vector routing (DVR) and link state routing (LSR). In the former one, routers do not have the knowledge of network topology, whereas in the later routers have this knowledge. In the present work, we have considered DVR-based networks. In [8-10] we have given a new and very important interpretation of the information present in the DVR tables of routers. This interpretation has resulted in the concept of pseudo-diameter.
Pseudo diameter of a router $r_i$ denoted as $P_d(r_i)$ is defined as follows.

$$P_d(r_i) = max \{c_{i,j}\}, \text{ where } c_{i,j} = cost(r_i, r_j), [1 \leq j \leq n, j \neq i] \text{ and } c_{i,j} \in DVR_i$$

In words, for a source router $r_i$, based on its DVR table, $DVR_i$, its pseudo diameter, denoted as $P_d(r_i)$, is the maximum value among the costs to reach from source $r_i$ to all other routers in a network. The implication of pseudo-diameter is that any other router is reachable from source router $r_i$ within the distance (i.e. cost / no. of hops) equal to the pseudo diameter of router $r_i$, $P_d(r_i)$. It thus directly relates to the physical location of router $r_i$. Pseudo diameter is not the actual diameter of the network, because it depends on the location of router $r_i$ in the network. So, different routers in the network may have different values for their respective pseudo diameters. Therefore, pseudo diameter $P_d$ is always less than or equal to the actual diameter of a network.

As an example, consider the network shown in Figure 1. We have shown the DVR tables of the routers. Note that the diameter of the network is 90. From router A's table, its pseudo diameter is 90, which is equal to the network diameter; whereas for router C it is 70 as is seen from C's DVR table. It means that if C is the source of a communication, the maximum cost to reach any other router will be 70, which is less than the network diameter of 90.

In this context, the following observations [10] are worth mentioning:

**Lemma 1**: *Let $S_i$ be the source and $d_i$, $d_j$, ....., $d_m$ be the respective reductions in the pseudo-diameter ($P_d$) of $S_i$ before a data packet arrives at its destination D, then $d_i + d_j + ......... + d_m \leq P_d$.*

**Lemma 2**: *Broadcasting algorithm based on DVR with pseudo diameter guarantees that each router in the network receives a copy of the packet sent by the broadcast source.*

## 3 Candidate Core Selection

We introduce a systematic approach to select in networks which use DVR for message candidate cores communication. This core selection is independent of any multicast group and will involve all routers in a network without having to learn about the entire network topology.
Let cost $c_{i,j}$ be measured in no. of hops between two routers. Let $r_i$ be any router in a network of $n$ routers and let $T_i$ denote the tree rooted at $r_i$ and $T_i(L)$ be its number of levels.
Therefore, $T_i(L) = P_d(r_i)$ = maximum cost to reach a leaf router $r_j$, because $c_{i,j} \leq P_d(r_i)$.
Hence, a router $r_k$ will be the first core to be selected,
   if $T_k(L) = min \{T_j(L)\}, \forall j$
   i.e. $P_d(r_k) = min \{P_d(r_j)\}, \forall j$
Note that even when the cost is measured differently, a router $r_k$ will be selected first in the candidate core set, if $P_d(r_k) = min \{P_d(r_j)\}, \forall j$.

A router $r_m$ is selected as the second core in the set,
   if $P_d(r_m) = min \{P_d(r_j)\}, j \neq k$

Similarly a third core (with the next lowest $P_d$ value) and so on, can be selected.

Note that during the core selection process in the event of a tie among routers, router with highest router ID (i.e. highest IP address) has a selection priority.

## 3.1 Algorithm Description

The following notations and data structures are used in the algorithm.

$n$ : total number of routers in a network;
*broad_message* $(P_d(r_i), r_i)$ : broadcast message from router $r_i$ with its pseudo-diameter $P_d$;
$dvr_i[][]$ : DVR table of router $r_i$, $[1 \leq i \leq n]$;
*core*[][] : two dimensional array that contains router IDs and corresponding pseudo diameters in ascending order of pseudo diameter;

**Procedure *broad*()** is called by each $r_i$ where $(1 \leq i \leq n)$ to broadcast $P_d(r_i)$. Each router controls the broadcast with its $P_d$ value, i.e. a router broadcasts message to its neighbor only if it is within $P_d$ range.

**Procedure receive_Broad($P_d(r_i)$, $r_i$)** is used by each router. This procedure has $P_d(r_i)$ and $r_i$ as input and generates *core*[][] that contains router IDs and their respective pseudo diameter values.

**Procedure Candidate_Core()** called by $r_i$, has *core*[][] as input. It sorts *core*[][] in ascending order of the $P_d$ values. If multiple routers have same $P_d$ value, the routers are sorted in the descending order of their IP addresses. At the end of this procedure, *core*[][] contains all routers sorted in order of increasing pseudo diameters. Based on a given application's needs, some of these routers starting from the beginning of the list may be considered as the candidate cores for multicore multicast by the same multicast group members.

**Algorithm Candidate_Core**

**Procedure broad()**
**Begin**
  *for each $r_i$ [$1 \leq i \leq n$]*
    $P_d(r_i) = max\ (dvr_i[j][2])\ [1 \leq j \leq n]$
    *send broad_message ($P_d\ (r_i)$, $r_i$)*
  *end for*
/* each node broadcasts its pseudo-diameter value ($P_d$ based broadcast)*/
**End**

**Procedure receive_Broad($P_d(r_i)$, $r_i$)**
**Begin**
  *for each $r_i$ [$1 \leq i \leq n$]*
    *for k = 1 to n*
      *if i ≠ k*
        $core[k][0]=r_k$
        $core[k][1]=P_d(r_k)$
      *end if*
    *end for*
  *end for*
/*core[][], a two-dimensional array that contains router ID and pseudo-diameter*/
  *Candidate_Core(core[][])*
**End**

**Procedure Candidate_Core(core[][])**
**Begin**
    *sort_asc(core[][])* /*sort in ascending order of $P_d$*/
    *if same $P_d$*
      *sort_desc(core[][])* /* in descending order */
    *end if*
  *end for*
**End**

## 3.2 An Example

In the present approach, it is observed that selection of the cores requires only the information about next hops and costs to reach all possible destinations; this information is present in the distance vector routing table of each router. Consider the example 8 router network as shown in Figure 1a. The DVR table of each router is shown in Figure 1b. Based on the DVR tables of the network, pseudo-diameter of all routers in the network can be obtained. Pseudo diameters of routers A, B, C, D, E, F, G, and H are 90, 90, 70, 80, 60, 90, 80, and 80 respectively.

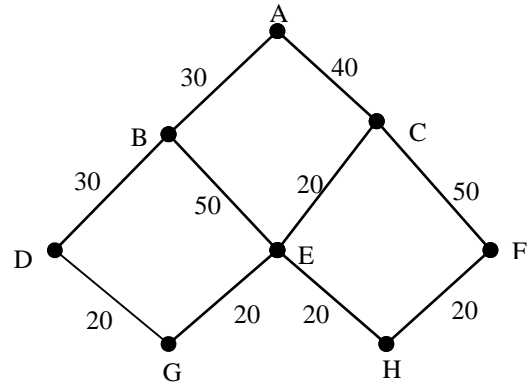

Figure 1a: An 8 Router network

| A | | |
|---|---|---|
| Dest. | Next | Delay |
| A | A | 0 |
| B | B | 30 |
| C | C | 40 |
| D | B | 60 |
| E | C | 60 |
| F | C | 90 |
| G | C | 80 |
| H | C | 80 |

| B | | |
|---|---|---|
| Dest. | Next | Delay |
| A | A | 30 |
| B | B | 0 |
| C | A | 70 |
| D | D | 30 |
| E | E | 50 |
| F | E | 90 |
| G | D | 50 |
| H | E | 70 |

| C | | |
|---|---|---|
| Dest. | Next | Delay |
| A | A | 40 |
| B | A | 70 |
| C | C | 0 |
| D | E | 60 |
| E | E | 20 |
| F | F | 50 |
| G | E | 40 |
| H | E | 40 |

| D | | |
|---|---|---|
| Dest. | Next | Delay |
| A | B | 60 |
| B | B | 30 |
| C | G | 60 |
| D | D | 0 |
| E | G | 40 |
| F | G | 80 |
| G | G | 20 |
| H | G | 60 |

| E | | |
|---|---|---|
| Dest. | Next | Delay |
| A | C | 60 |
| B | B | 50 |
| C | C | 20 |
| D | G | 40 |
| E | E | 0 |
| F | H | 40 |
| G | G | 20 |
| H | H | 20 |

| F | | |
|---|---|---|
| Dest. | Next | Delay |
| A | C | 90 |
| B | H | 90 |
| C | C | 50 |
| D | H | 80 |
| E | H | 40 |
| F | F | 0 |
| G | H | 60 |
| H | H | 20 |

| G | | |
|---|---|---|
| Dest. | Next | Delay |
| A | E | 80 |
| B | D | 50 |
| C | E | 40 |
| D | D | 20 |
| E | E | 20 |
| F | E | 60 |
| G | G | 0 |
| H | E | 40 |

| H | | |
|---|---|---|
| Dest. | Next | Delay |
| A | E | 80 |
| B | E | 70 |
| C | E | 40 |
| D | E | 60 |
| E | E | 20 |
| F | F | 20 |
| G | E | 40 |
| H | H | 0 |

Figure 1b: DVR table of each router

Each router $r_i$ broadcasts its $P_d$ value to all other routers in the network ($r_i$, $1 \leq i \leq n$). At the end of the broadcast, each router contains $P_d$ of every other router in the network as shown in Figure 2a. Observe in Figure 2a that there is more than one router with same $P_d$ value. Routers D, G, and H have the same $P_d$ value, viz., 80. If this situation arises, the routers are sorted in descending order of their router IDs. Without any loss of generality assume that H's ID is the highest among the three. This means that router H has the selection priority over the other two routers. Router $r_i$ calls **procedure Candidate_Core** which returns the sorted *core*[][] array as shown in Figure 2b. Let us assume that we select only three candidate cores.

So the candidate core set is {E(60), C(70), H(80)} where the respective $P_d$'s have been bracketed.

| $r_i$ | $P_d(r_i)$ |
|---|---|
| A | 90 |
| B | 90 |
| C | 70 |
| D | 80 |
| E | 60 |
| F | 90 |
| G | 80 |
| H | 80 |

Figure 2a: *core*[][] before Candidate_Core procedure call

| $r_i$ | $P_d(r_i)$ |
|---|---|
| E | 60 |
| C | 70 |
| H | 80 |
| G | 80 |
| D | 80 |
| F | 90 |
| B | 90 |
| A | 90 |

Figure 2b: Sorted *core*[][] after Candidate_core procedure call

### 3.3 Performance

The complexity of candidate core selection approach is $O(n^2)$, where $n$ is the total number of nodes in the network. In our approach a router does not have the complete topological information. The proposed approach is compared with some important existing core selection approaches. Complexities of these existing methods are briefly discussed below.

Maximum path count (MPC) core selection method [3] finds the shortest paths for all pairs of nodes in the given network. Complexity of this approach is $O(n^2)$ where $n$ is the number of nodes in the network. Minimum average distance (MAD) method [3] finds the average distance along the shortest paths from each node to all other nodes in the network. The nodes are sorted in ascending order of their average distance. The first few nodes are selected to be the candidate cores. Complexity of this approach is $O(n^2)$, where $n$ is the total number of nodes in the network. However, in both MPC and MAD a router needs to have the complete topological information unlike our approach.

In Delay Variant Multicast Algorithm (DVMA) [5] the worst case complexity of DVMA is $O(klmn^4)$, where $k$, $l$ are the numbers of path generated, $m$ is the size of the multicast group, and $n$ is the number of nodes in the network. OptTree [3] method suggests an optimization criterion whose complexity is $O(|M|^3|C|)$, where $M$ is the number of multicast group members and $C$ is the number of candidate cores.

## 4 Locality-Based Multicore Multicasting

It is understood that every sender, be it a multicast group member or not, wants its multicast traffic to arrive at the group members as fast as possible; in other words the end-to-end delay of the multicast traffic should be as low as possible from the viewpoint of a sender. We will now state a method of how during a multicast session a sender node will determine which core will give the minimum end-to-end delay considering the physical location of itself and those of the candidate cores. We use the following notations.

A candidate core set $S_c$ is denoted as:

$S_c = \{Core_k\}$, $1 \leq k \leq m$, and $P_d(Core_1) \leq ... \leq P_d(Core_m)$.

End-to end delay $ED_i^k$ between a router $r_i$ and a core $Core_k$ is denoted as:

$ED_i^k = [P_d(Core_k) + C_{ik}]$, where $C_{ik}$ is the cost between router $r_i$ and $Core_k$ as is present in the DVR table of router $r_i$.

A sender router $r_i$ selects the $Core_j$ for which:

$$ED_i^j = min\ \{ED_i^k\},\ 1 \leq k \leq m$$

**Example**: Consider the example of Figure 1. Let router A be a sender. We have seen that the candidate core set is
$Sc = \{E, C, H\}$.
We note that the costs from A to the cores E, C, and H are: 60, 40, and 80 respectively.
Router A computes $ED_A^E = 120$; $ED_A^C = 110$; $ED_A^H = 160$. Therefore, router A selects core C for its multicast traffic delivery.

Salient features of the proposed approach are as follows:
1. Information needed is only the routing tables of the routers.
2. Each router independently decides which routers should be the candidate cores and this decision is unanimous.
3. Each sender independently decides which core to choose for sending traffic, from the viewpoint of its locality as well as the locality of the chosen core.
4. The proposed approach is independent of any central control unlike some other related works [15].

## 5 Conclusion

In this paper, we have presented a candidate core selection approach for networks that use DVR as the unicast routing protocol. A sender selects a core from the candidate set to satisfy its end-to-end delay needs. So different senders sending traffic to the same multicast group may use different cores; thereby resulting in a possible overall faster delivery of the multicast traffic, while at the same time resulting in a possible even load distribution on multiple cores.